\documentclass[12pt]{article}
\usepackage[dvips]{color}
\usepackage{epsfig}
\usepackage{amsmath}
\usepackage{graphicx}
\usepackage{cite}
\usepackage{subcaption}

\begin{document}

\title{\bf Modified Friedmann equations from DSR-GUP}
\author{{ \"{O}zg\"{u}r \"{O}kc\"{u} \thanks{Email: ozgur.okcu@ogr.iu.edu.tr}\hspace{1mm}, Christian Corda \thanks{Email: cordac.galilei@gmail.com},
Ekrem Aydiner \thanks{Email: ekrem.aydiner@istanbul.edu.tr}} \\
{\small {\em Department of
		Physics, Faculty of Science, Istanbul University,
		}}\\{\small {\em Istanbul, 34134, Turkey}} }
\maketitle

\begin{abstract}
Considering the modified entropy-area relation from DSR-GUP (Doubly special relativity-Generalized uncertainity principle), we obtain the modified Friedmann equations from the first law of thermodynamics at apparent horizon. Due to the importance of GUP at Planck scale, we investigate the Friedmann equations and show the maximum energy density $\rho$ at Planck scale. Since GUP implies a minimal length, we find a minimum apparent horizon which has a potential to remove the Big Bang singularity. Furthermore, we analyse the effects of DSR-GUP on deceleration parameter $q$ for the equation of state $p=\omega\rho$ and the flat case. Finally, we check the validity of the generalized second law (GSL) of thermodynamics and show that it is valid for all eras of the universe for any spatial curvature.  \\

{\bf Keywords:} Cosmology; DSR-GUP; generalized uncertainty principle.
\end{abstract}
\begin{quote}\textbf{The Authors dedicate this paper to people who died due to Covid-19 and those who fight it
\newpage{}}
\end{quote}

\section{Introduction}

Considering black holes as the thermodynamic systems reveals the fundamental connection between the gravity and laws of thermodynamics \cite{Bekenstein1972,Bekenstein1973,Bardeen1973,Hawking1974,Bekenstein1974,Hawking1975}. A black hole has temperature and entropy proportional to its surface gravity and horizon area, respectively. Motivated by the thermodynamics properties of black hole, Jacobson \cite{Jacobson1995} first interpreted the Einstein field equation as an equation of state. He derived the field equation from the proportionality of entropy and horizon area together with Clausius relation  $\delta Q=TdS$ which connects the heat, temperature and entropy. Later, the deeper fundamental connection between gravitational dynamics and horizon thermodynamics was indicated in many papers \cite{Padmanabhan2002,Eling2006,Cai2005,Akbar2006,Akbar2007,Cai2007a,Cai2007b,Cai2008,Sheykhi2010a,Awad2014,Salah2017,Nojiri2006,Sheykhi2010b,Sheykhi2007a,Sheykhi2007b,Sheykhi2009,Sheykhi2018,Karami2011}. Motivated by Jacobson's work, Cai and Kim \cite{Cai2005} obtained the $(n+1)$-dimensional Friedmann equations from the first law of thermodynamics ($-dE=T_{h}dS_{h}$) at apparent horizon. Here $-dE$ is interpreted as the amount of energy flux crossing the apparent horizon for the infinitesimal time interval at fixed horizon radius. We assume that the apparent horizon has temperature and entropy given as follows:
\begin{equation}
\label{Temp-entropy1}
T_{h}=\frac{1}{2\pi\tilde{r_{A}}},\qquad\qquad\qquad S_{h}=\frac{A}{4},
\end{equation}
where $A$ is the area of the apparent horizon\footnote{We use the units $\hbar=c=G_{N}=l_{Pl}^{2}=1$}. Using the corresponding entropy-area formula in the Gauss-Bonnet and Lovelock gravities, they were derived the Friedmann equations in each gravity. Following Cai and Kim's approach, Friedmann equations in the scalar$-$tensor and $f(R)$ gravities were obtained in Ref.\cite{Akbar2006} \footnote{We recall that the extended gravity framework arises from the necessity to extend standard general relativity in order to attempt to achieve the famous Dark Energy and Dark Matter problems \cite{Odintsov2011}. In this framework, the recent starting of the gravitational wave astronomy with the famous detections of LIGO \cite{Abbott2017} could be, in principle, decisive to confirm the physical consistence of  standard general relativity, or, alternatively, to endorse the framework of extended theories of gravity \cite{Corda2009}. In fact, some differences between general relativity and alternative theories can be pointed out in the linearized theory of gravity through different interferometer response functions \cite{Corda2009}.}. Despite one can obtain the Friedmann equations in \cite{Cai2005}, it can be seen that the temperature of apparent horizon is not proportional to surface gravity $\kappa$ of the horizon, since surface gravity is given by
\begin{equation}
\label{kappa}
\kappa=-\frac{1}{\tilde{r_{A}}}\left(1-\frac{\dot{\tilde{r_{A}}}}{2H\tilde{r_{A}}}\right),
\end{equation}
where dot denotes the derivative with respect to time and $H$ is Hubble parameter. Since the $\tilde{r_{A}}$ does not change, the first law of thermodynamics, which is proposed in Ref.\cite{Cai2005}, is satisfied. However, this approximation has a limitation on the equation of state as $p\approx-\rho$, i.e., equation of state implies the vacuum energy or de Sitter spacetime. Assuming the temperature $T_{h}=\frac{\kappa}{2\pi}$ and entropy $S_{h}=\frac{A}{4}$, Akbar and Cai \cite{Akbar2007} showed that the differential form of Friedmann equations can be rewritten as the first law of thermodynamics at apparent horizon
\begin{equation}
\label{firstLaw}
dE=T_{h}dS_{h}+WdV,
\end{equation}
where $W$ is the work density which is given in terms of energy density $\rho$ and pressure $p$ of the matter in the universe, $E=\rho V$ is the total energy inside the apparent horizon, and $V$ is the volume of the apparent horizon.

On the other hand, it is well known that entropy-area relation can be modified via various quantum gravity (QG) approaches \cite{Govindarajan2001,Mann1998,Sen2013} since the QG effects are remarkable at the Planck scale. For example, motivated by loop quantum gravity (LQG), a modified version of Friedmann equations was obtained in Refs.\cite{Cai2008,Sheykhi2010a}. Besides LQG, one can consider modification of entropy-area relation from GUP which is one of the phenomenological QG model and a modification of the standard uncertainty principle\cite{Maggiore1993,Kempf1995,Nozari2012}.  One of the most characteristic implications in GUP is the concept of minimal length.  Since the minimal length notion may cure the singularities in general relativity, it is also interesting to consider both applications of black hole thermodynamics and cosmology in the context of the GUP \cite{Adler2001,Nozari2005,Nozari2008,Nowakowski2009,Arraut2009,Banerjee2010,Nozari2012b,Ali2012,Gangopadhyay2014,Abbasvandi2016,Feng2016,Xiang2009,Sun2018,Okcu2019,Zhu2009,Chen2005}. Taking into account the simplest form of GUP, Awad and Ali \cite{Awad2014} obtained the modified Friedmann equations. They showed that modified Friedmann equations exhibit the maximum energy denisty (order of Planck energy density) at minimal length. Their work extended in Ref. \cite{Salah2017} for a new version of GUP. Similarly, an upper bound was shown for the energy of universe. Furthermore, it is possible to define a cyclic universe from the modified GUP.  Another modification of standart uncertainty principle may be considered in the context of DSR. We recall that DSR modifies standard special relativity by adding an observer-independent maximum energy scale and minimum length scale (the Planck energy and Planck length) to the observer-independent maximum velocity (the speed of light)of standard special relativity \cite{Amelino2010}.
In Refs. \cite{Chung2018,Maghsoodi2019}, authors considered the GUP based on DSR and renamed it as DSR-GUP. Apart from special relativity, DSR also has an extra upper bound as Planck energy. Therefore, DSR plays a crucial role in investigating the quantum gravity effects near the Planck scale. Motivated by this, we would like to investigate the modified Friedmann equations for the DSR-GUP in the present paper.

The paper is organized as follows: In the next section, we review the DSR-GUP and obtain the modified entropy-area relation. In the third section, we obtain the modified Friedmann equation from DSR-GUP. In the fourth section, we investigate the effects of DSR-GUP on deceleration parameter. In the fifth section, we check the validity of GSL for the modified Friedmann equations. Finally, the conclusions are presented in the last section.

 \section{GUP based on DSR}
 In this section, we briefly review the DSR-GUP and calculate modified entropy-area relation \cite{Chung2018,Maghsoodi2019}. Let us start to give the new form of GUP which is based on DSR
 \begin{equation}
 \label{DSR-GUP}
 \Delta x\Delta p\geq\frac{1}{2}\left(1-2\alpha+\frac{\Delta p^{2}}{\epsilon_{p}^{2}}\right)
 \end{equation}
 where $\epsilon_{p}$ is Planck energy and $0<\alpha<1/2$. In order to obtain the modified entropy-area relation, we consider the heuristic analysis in Ref.\cite{Xiang2009}. Firstly, we need to solve the inequality of DSR-GUP for the lower bound of $\Delta p$
 \begin{equation}
 \label{lowerBound}
 \Delta p\geq\Delta x\epsilon_{p}^{2}-\epsilon_{p}\sqrt{\Delta x^{2}\epsilon_{p}^{2}+2\alpha-1},
 \end{equation}
 which gives the standart uncertainty when $\alpha\rightarrow0$ and $\epsilon_{p}\rightarrow\infty$. Using Eq.(\ref{lowerBound}) with $\Delta x \sim 2r_{h}$ event horizon, one can write $\Delta x\Delta p$
 \begin{equation}
 \label{newDxDp}
 \Delta x\Delta p\geq4r_{h}^{2}\epsilon_{p}^{2}-2r_{h}\epsilon_{p}\sqrt{4r_{h}^{2}\epsilon_{p}^{2}+2\alpha-1}.
 \end{equation}
 Moreover, the smallest increase of the black hole area can be considered as
 \begin{equation}
 \label{area2}
 \Delta A\geq\Delta x\Delta p.
 \end{equation}
 when a particle is absorbed by a black hole. Therefore, we obtain the increase of area
 \begin{equation}
 \label{AreaDSRGUP}
 \Delta x\Delta p\geq4r_{h}^{2}\epsilon_{p}^{2}\gamma-2r_{h}\epsilon_{p}\gamma\sqrt{4r_{h}^{2}\epsilon_{p}^{2}+2\alpha-1},
 \end{equation}
 where $\gamma$ is the calibration factor. We know that minimum increase of entropy $(\Delta S)_{min}=\ln2$ in the information theory. Hence, we obtain the modified area-entropy relation based on DSR-GUP \cite{Maghsoodi2019}
 \begin{equation}
 \label{areaEntropy}
 \frac{dS}{dA}\approx\frac{(\Delta S)_{min}}{(\Delta A)_{min}}=\frac{1}{32\epsilon_{p}^{2}r_{h}^{2}\left[1-\sqrt{1+\frac{2\alpha-1}{4\epsilon_{p}^{2}\gamma r_{h}^{2}\left(1-\sqrt{1+\frac{2\alpha-1}{4\epsilon_{p}^{2}r_{h}^{2}}}\right)}}\right]}
 \end{equation}
 where the calibration factor is obtained as $\gamma=8\ln2$ in the limits of $\alpha\rightarrow0$ and $\epsilon_{p}\rightarrow\infty$.
 \section{Modified Friedmann equations}
 We firstly start to review the basic elements of Friedmann-Robertson-Walker (FRW) universe. The line element of FRW universe is given by
 \begin{equation}
 \label{lineElement}
 ds^{2}=h_{ab}dx^{a}dx^{b}+\tilde{r}d\Omega^{2},
 \end{equation}
 where $\tilde{r}=a(t)r$, $a$ is the scale factor, $x^{a}=(t,r)$, $h_{ab}=diag\left(-1,a^{2}/(1-kr^{2})\right)$ is two dimensional metric, and $k$ corresponds to the values $-1$, $0$, $1$ for the open, flat and closed universe, respectively. The dynamical apparent horizon is given by
 \begin{equation}
 \label{apparentHor}
 \tilde{r_{A}}=ar=\frac{1}{\sqrt{H^{2}+k/a^{2}}},
 \end{equation}
 where $H=\dot{a}/a$ is the Hubble parameter. The temperature of the apparent horizon is obtained from Eq.(\ref{kappa})
 \begin{equation}
 \label{Temp}
 T_{h}=\frac{\kappa}{2\pi}=-\frac{1}{2\pi\tilde{r_{A}}}\left(1-\frac{\dot{\tilde{r_{A}}}}{2H\tilde{r_{A}}}\right).
 \end{equation}
 Since entropy is the function of area, we can write the general expression for entropy in the following form:
 \begin{equation}
 \label{entropy}
 S_{h}=\frac{f(A)}{4},
 \end{equation}
 and the differential of entropy is given by
 \begin{equation}
 \label{difentropy}
 \frac{dS_{h}}{dA}=\frac{f'(A)}{4},
 \end{equation}
 where prime denotes the derivative with respect to area $A=4\pi\tilde{r_{A}}^{2}$. By assuming the the matter and energy of the universe as an ideal fluid, so energy-momentum tensor yields
 \begin{equation}
 \label{energyMomentumTensor}
 T_{\mu\nu}=(\rho+p)u_{\mu}u_{\nu}+pg_{\mu\nu}
 \end{equation}
 where $u_{\mu}$ is the four velocity of fluid. The conservation of energy-momentum tensor, i.e., $T_{;\nu}^{\mu\nu}=0$ leads to the continuity equation as
 \begin{equation}
 \label{continuityEqu}
 \dot{\rho}+3H(\rho+p)=0.
 \end{equation}
 Following the arguments of Ref.\cite{Hayward1998}, we can define the work density as
 \begin{equation}
 \label{workDensity}
 W=-\frac{1}{2}T^{ab}h_{ab}=\frac{1}{2}(\rho-p).
 \end{equation}
 Work density $W$ corresponds to work done by the volume change of universe. Now, we can calculate the terms of Eq. (\ref{firstLaw}). Since the volume and the total energy of universe are $V=\frac{4}{3}\pi \tilde{r_{A}}^{3}$ and $E=\rho V$, differential of $E$ is given by using the continuity equations in Eq.(\ref{continuityEqu})
 \begin{equation}
 \label{dE}
 dE=\rho dV+Vd\rho=4\pi\rho\tilde{r_{A}}^{2}d\tilde{r_{A}}-4\pi(\rho+p)\tilde{r_{A}}^{3}Hdt,
 \end{equation} 
 and $WdV$ term is calculated as
 \begin{equation}
 \label{WdV}
 WdV=2\pi(\rho-p)\tilde{r_{A}}^{2}d\tilde{r_{A}}.
 \end{equation}
 Finally, $TdS$ term is found
 \begin{equation}
 \label{TdS}
 T_{h}dS_{h}=-\left(1-\frac{\dot{\tilde{r_{A}}}}{2H\tilde{r_{A}}}\right)f'(A)d\tilde{r_{A}}.
 \end{equation}
 Combining Eqs. (\ref{dE}), (\ref{WdV}) and (\ref{TdS}) in the first law at the apparent horizon and using the relation
 \begin{equation}
 \label{difAppa}
 d\tilde{r_{A}}=-H\tilde{r_{A}}^{3}\left(\dot{H}-\frac{k}{a^{2}}\right)dt,
 \end{equation}
 we get
 \begin{equation}
 \label{intEq}
 \frac{f'(A)}{\tilde{r_{A}}^{3}}\tilde{dr_{A}}=4\pi(\rho+p)Hdt.
 \end{equation}
 If we use the continuity equation in Eq.(\ref{continuityEqu}) with the above equation, we get the differential form of Friedmann equation as
 \begin{equation}
 \label{diffFriedman}
 \frac{f'(A)}{\tilde{r_{A}}^{3}}d\tilde{r_{A}}=-\frac{4\pi}{3}d\rho,
 \end{equation}
 and rearranging the Eqs. (\ref{difAppa}) and (\ref{intEq}), one can simply find the dynamical equation as follows:
 \begin{equation}
 \label{dynamicalEquation}
 f'(A)\left(\dot{H}-\frac{k}{a^{2}}\right)=-4\pi(\rho+p).
 \end{equation}
 Using the modified entropy-area relation in Eq.(\ref{areaEntropy}) with Eq.(\ref{difentropy}), then we can find $f'(A)$. Therefore, we find the Friedmann equations from Eqs.(\ref{diffFriedman}) and (\ref{dynamicalEquation})
 \begin{equation}
 \label{firstFried}
 -\frac{4\pi}{3}\rho=-\frac{1}{4\tilde{r_{A}}^{2}(1-2\alpha)}+\frac{(4\tilde{r_{A}}^{2}\epsilon_{p}^{2}+2\alpha-1)^{3/2}}{12\tilde{r_{A}}^{3}\epsilon_{p}(1-2\alpha)^{2}}+C,
 \end{equation}
 \begin{equation}
 \label{secondFried}
 \left(\dot{H}-\frac{k}{a^{2}}\right)\frac{1}{8\epsilon_{p}^{2}\left(1-\sqrt{1+\frac{2\alpha-1}{4\epsilon_{p}^{2}\tilde{r_{A}}^{2}}}\right)\tilde{r_{A}}^{2}}=4\pi(\rho+p),
 \end{equation}
 where $C$ is the integration constant and can be determined from initial conditions. As the universe expands, $\tilde{r_{A}}$ goes to infinity and energy density is the vacuum energy $\rho_{vac}=\Lambda$, where $\Lambda$ is the cosmological constant. So we find the integration constant $C=-\frac{4\pi\Lambda}{3}-\frac{2\epsilon_{p}^{2}}{3(1-2\alpha)^{2}}$ and first Friedmann equation is given
 \begin{equation}
 \label{firstFriedmannEqu}
 \frac{8\pi}{3}(\rho-\Lambda)\left(1-2\alpha\right)^{2}=\frac{1}{6}\left[\frac{3(1-2\alpha)}{\tilde{r_{A}}}+8\epsilon_{p}^{2}\left(1-\left(1+\frac{2\alpha-1}{4\tilde{r_{A}}^{2}\epsilon_{p}^{2}}\right)^{\frac{3}{2}}\right)\right].
 \end{equation}
 Note that the modified Friedmann equations can be reduced to the standard forms in the limits of $\alpha\rightarrow0$ and $\epsilon_{p}\rightarrow\infty$.
 
 Now, let us investigate the Eq.(\ref{firstFriedmannEqu}). In order to get a real and positive energy density $\rho$, $\tilde{r}_{A}^{min}$ should has a minimum value which is given by
 \begin{equation}
 \label{minimumLenght}
 \tilde{r}_{A}^{min}=\frac{\sqrt{1-2\alpha}}{2\epsilon_{p}},
 \end{equation}
 which removes the singularity at the beginning. We stress that such a removal of the initial singularity depends on the quantum behavior of the GUP of Eq.(\ref{DSR-GUP}). The GUP expressed by Eq.(\ref{DSR-GUP}) implies indeed a non-zero lower bound on the minimum value of the uncertainty on the particles' position $\left(\Delta x\right)$ which is of order of the Planck length. This issue has no classical correspondence because in standard general relativity all time-like radial geodesics of the particles in the Universe start at the "initial point" $\tilde{r_{A}}=0$  and it is impossible to extend the global space-time manifold beyond that "initial point". This is the meaning of the classical initial singularity.
 At the minimum apparent horizon, we can obtain the maximum allowed value of energy density as $\rho_{max}=\frac{5\epsilon_{P}^{2}}{4\pi(1-2\alpha)^{2}}+\Lambda$ for the inflationary scale. Since $\Lambda$ is too small, it can be neglected and we can easily see that energy density is order of Planck energy. Apart from standard Friedmann equations, we find a non-zero minimum apparent horizon and finite maximum energy density for the modified Friedmann equations. It is also clear that $\tilde{r}_{A}^{min}$ goes to zero and $\rho_{max}$ diverges in the limits of $\alpha\rightarrow0$ and $\epsilon_{p}\rightarrow\infty$.
 
 Using the Eq.(\ref{apparentHor}), we give the Friedmann equation in term of Hubble parameter $H$ as follows:
 	\begin{equation}
 	\label{firstFriedmann2}
 	\frac{8\pi}{3}(\rho-\Lambda)(1-2\alpha)^{2}=\frac{1}{6}\left[3(1-2\alpha)\left(H^{2}+\frac{k}{a^{2}}\right)+8\epsilon_{p}^{2}\left(1-\left(1+\frac{(2\alpha-1)}{4\epsilon_{p}^{2}}\left(H^{2}+\frac{k}{a^{2}}\right)\right)^{3/2}\right)\right],
 	\end{equation}
 \begin{equation}
 \label{secondFriedmann2}
 \left(\dot{H}-\frac{k}{a^{2}}\right)\frac{1}{8\epsilon_{p}^{2}}\frac{\left(H^{2}+\frac{k}{a^{2}}\right)}{1-\sqrt{1+\frac{(2\alpha-1)}{4\epsilon_{p}^{2}}\left(H^{2}+\frac{k}{a^{2}}\right)}}=-4\pi(\rho+p),
 \end{equation}
 which will be used to investigate the deceleration parameter in the next section.
 \section{Deceleration parameter}
 In this section, we investigate the effects of DSR-GUP on the deceleration parameter which is defined as
 \begin{equation}
 \label{decelerationParameter}
 q=-1-\frac{\dot{H}}{H^2},
 \end{equation}
 where the positive $q$ implies deceleration, while negative q implies the acceleration. Now, choosing the equation of state as $p=\omega \rho$, and combining Eqs. (\ref{firstFriedmann2}) and (\ref{secondFriedmann2}) with Eq.(\ref{decelerationParameter}) gives the deceleration parameter as
 	\begin{eqnarray}
 	\label{dP}
 	&q=-1+\frac{2(\omega+1)\epsilon_{p}^{2}}{(1-2\alpha)^{2}H^{2}}&\\&\nonumber\times\left[3(1-2\alpha)+\frac{8\epsilon_{p}^{2}}{H^{2}}\left(1-\left(1+\frac{\left(2\alpha-1\right)}{4\epsilon_{p}^{2}}H^{2}\right)^{3/2}\right)+\frac{16\pi(1-2\alpha)^{2}\Lambda}{H^{2}}\right]\times\left[1-\sqrt{1+\frac{2\alpha-1}{4\epsilon_{p}^{2}}H^{2}}\right],
 	\end{eqnarray}
 for the flat case $k=0$. We choose the Euclidean case for the shape of the Universe because it seems in agreement with current cosmological observations  \cite{Odintsov2011}. Except the vacuum dominated universe ($p=-\rho$), we can mostly neglect the cosmological constant $\Lambda$. First, we calculate the value of deceleration parameter for the inflationary stage. At the minimal length, we can find the maximum H as
 \begin{equation}
 \label{Hmax}
 H_{max}=\frac{2\epsilon_{p}}{\sqrt{1-2\alpha}}
 \end{equation}
 which leads to
 \begin{equation}
 \label{qAtPlanck}
 q=\frac{3}{2}+\frac{5\omega}{2}
 \end{equation}
 for the inflationary stage. Interestingly, we find that $\omega$ should satisfy the condition $\omega<-3/5$ to imply the accelerated universe for the beginning of the inflationary stage.  Since effects of GUP may be sufficiently small for radiation and matter dominated eras, we can expanded the deceleration parameter as follows:
 \begin{equation}
 \label{qSeries}
 q=\frac{1}{2}(1+3\omega)+\frac{3H^{2}(1+\omega)(1-2\alpha)}{64\epsilon_{p}^{2}}+... .
 \end{equation}
 From Eq.(\ref{qSeries}), deceleration parameter for the radiation ($\omega=1/3$) and the matter ($\omega=0$) dominated eras can be given by
 \begin{equation}
 \label{radq}
 q=1+\frac{(1-2\alpha)H^{2}}{16\epsilon_{p}^{2}},
 \end{equation}
 \begin{equation}
 \label{matq}
 q=\frac{1}{2}+\frac{3H^{2}(1-2\alpha)}{64\epsilon_{p}^{2}},
 \end{equation}
 respectively. Remind that $0<\alpha<1/2$, correction term certainly gives the positive contribution to both cases. So taking into account effects of DSR-GUP, we can deduce that expansion of universe  is more decelerated for the radiation and matter dominated eras. Furthermore, DSR-GUP effects are not an alternative to dark energy (DE) for the matter dominated universe since Eq.(\ref{matq}) is always positive\footnote{In contrast to our results, an accelerated universe is possible without the invoking DE. The reader may refer to Refs. \cite{Sheykhi2018} and \cite{Sefiedgar2017} which consider Tsallis and  rainbow gravity corrections to Friedmann equations, respectively.}. Since the correction term in Eq.(\ref{qSeries}) is too small for the late time, one may consider that most contribution comes from the first term. Eq.(\ref{qSeries}) implies that it must be $\omega<-1/3$ to explain the acceleration of late time universe. Therefore, we still need DE for the late time of universe.
 \section{Generalized second law of thermodynamics}
 In this section, we want to check the validity of the GSL in the presence of DSR-GUP. According to the GSL of thermodynamics, total entropies of matter fields and apparent horizon do not decrease with time. Rearranging the Eq.(\ref{intEq}), one can obtain
 \begin{equation}
 \label{intEq2}
 \dot{\tilde{r_{A}}}=32\pi(\rho+p)\tilde{r_{A}}^{5}H\epsilon_{p}^{2}\left(1-\sqrt{1+\frac{2\alpha-1}{4\tilde{r_{A}}^{2}\epsilon_{p}^{2}}}\right).
 \end{equation}
 It is clear that the sign of Eq.(\ref{intEq2}) depends on the sign of $\rho+p$. Combining the Eqs.(\ref{intEq2}) and (\ref{TdS}),
 we obtain the following expression as
 	\begin{equation}
 	\label{GSL}
 	T_{h}\dot{S_{h}}=4\pi(\rho+p)\tilde{r_{A}}^{3}H\left[1-16\pi(\rho+p)\epsilon_{p}^{2}\tilde{r_{A}}^{4}\left(1-\sqrt{1+\frac{2\alpha-1}{4\tilde{r_{A}}^{2}\epsilon_{p}^{2}}}\right)\right],
 	\end{equation}
 which may violate the second law of thermodynamics for accelerated universe. Therefore, we should check the validity of GSL of thermodynamics.
 
 Now, let us consider the Gibbs equation \cite{Izquierdo2006} which is defined as
 \begin{equation}
 \label{GE}
 T_{m}dS_{m}=d(\rho V)+pdV=Vd\rho+(\rho+p)dV,
 \end{equation}
 where $T_{m}$ and $S_{m}$ are the temperature and the entropy of the matter fields inside the horizon. We use the assumption that the apparent horizon remains in equilibrium with the system. So we have $T_{m}=T_{h}$. Form Eq.(\ref{GE}), we can find 
 	\begin{equation}
 	\label{GSL2}
 	T_{h}\dot{S_{m}}=-4\pi(\rho+p)\tilde{r_{A}}^{3}H\left[1-32\pi(\rho+p)\epsilon_{p}^{2}\tilde{r_{A}}^{4}\left(1-\sqrt{1+\frac{2\alpha-1}{4\tilde{r_{A}}^{2}\epsilon_{p}^{2}}}\right)\right].
 	\end{equation}
 From Eqs. (\ref{GSL}) and (\ref{GSL2}), we can obtain the time evolution of entropy
 \begin{equation}
 \label{GSLT}
 T_{h}(\dot{S_{h}}+\dot{S_{m}})=64\pi^{2}(\rho+p)^{2}H\epsilon_{p}^{2}\tilde{r_{A}}^{7}\left(1-\sqrt{1+\frac{2\alpha-1}{4\tilde{r_{A}}^{2}\epsilon_{p}^{2}}}\right)
 \end{equation}
 which is useful to check the GSL of thermodynamics. The right hand side of the above equation is always a non-decreasing for the all eras of the universe. Therefore, the GSL of thermodynamics is always valid for the all eras of the universe.
 \section{Conclusions}
 In this work, we obtained the DSR-GUP modified Friedmann equations from the first law of thermodynamics at apparent horizon. We investigated the modified Friedmann equations since GUP effects are not negligible at Planck scale.  We showed a nonzero minimum apparent horizon which leads to a maximum and finite energy density. These may have the potential to remove the singularity at beginning of universe. Moreover, we studied the DSR-GUP effects on the deceleration parameter for the flat case and equation of state as $p=\omega \rho$. We found that acceleration at the beginning of inflation imply the condition $\omega<-3/5$. The effects of DSR-GUP were also investigated for the radiation and matter dominated eras. The deceleration parameter is still positive in the presence of DSR-GUP. DSR-GUP may have contribution to deceleration for the radiation and matter dominated eras. Hence, DSR-GUP effects are not an alternative to DE for the matter dominated universe. As for the late time, we still need DE to explain the accelerated late time expansion. Finally, we checked the validity of GSL of thermodynamics and found that it is always valid for the all eras of universe.
 
In contrast to standard Friedmann equations, the DSR-GUP modified Friedmann equations have more reasonable and acceptable properties such as minimal length and maximum energy density. 

For the sake of completeness, we take the opportunity to cite some further important work on the GUP\cite{Scardigli,Scardigli2,Scardigli3,Scardigli4}. In particular, in \cite{Scardigli3} the Authors introduced  modification of the background metric due to GUP, while in \cite{Scardigli4} interesting cosmological consequences of the GUP have  been discussed.

\end{document}